\documentclass[10pt,a4paper,twocolumn,english,prl,aps,showpacs,floatfix,groupedaddress,superscriptaddress]{revtex4}
\usepackage[]{fontenc}
\usepackage[latin9]{inputenc}
\usepackage{color}
\usepackage{verbatim}
\usepackage{amsmath}
\usepackage{graphicx}
\usepackage{amssymb}
\usepackage{esint}

\makeatletter

\providecommand{\tabularnewline}{\\}
\newcommand{\lyxdot}{.}

\@ifundefined{textcolor}{}
{%
 \definecolor{BLACK}{gray}{0}
 \definecolor{WHITE}{gray}{1}
 \definecolor{RED}{rgb}{1,0,0}
 \definecolor{GREEN}{rgb}{0,1,0}
 \definecolor{BLUE}{rgb}{0,0,1}
 \definecolor{CYAN}{cmyk}{1,0,0,0}
 \definecolor{MAGENTA}{cmyk}{0,1,0,0}
 \definecolor{YELLOW}{cmyk}{0,0,1,0}
 }

\@ifundefined{definecolor}
 {\usepackage{color}}{}
\@ifundefined{definecolor}{\@ifundefined{definecolor}
 {\usepackage{color}}{}
}{}\@ifundefined{definecolor}{\@ifundefined{definecolor}{\@ifundefined{definecolor}
 {\usepackage{color}}{}
}{}}{}\def\1{1\negthickspace{\rm I}}

\makeatother

\usepackage{babel}

\makeatother

\usepackage{babel}

\begin{document}

\title{Hidden XY structure of the bond-charge Hubbard model}

\author{Marco Roncaglia}

\affiliation{Dipartimento di Fisica del Politecnico, corso Duca degli Abruzzi
24, I-10129 Torino, Italy}

\author{Cristian Degli Esposti Boschi}

\altaffiliation{On leave from CNISM, Unità di Ricerca del Dip. di Fisica dell'Università di Bologna, where this work was started.}

\affiliation{CNR-IMM, Sezione di Bologna, via Gobetti 101, I-40129 Bologna, Italy}

\author{Arianna Montorsi}

\affiliation{Dipartimento di Fisica del Politecnico, corso Duca degli Abruzzi
24, I-10129 Torino, Italy}

\date{\today}
\begin{abstract}
The repulsive one-dimensional Hubbard model with bond-charge interaction
(HBC) in the superconducting regime is mapped onto the spin-$1/2$
XY model with transverse field. We calculate correlations and phase
boundaries, realizing an excellent agreement with numerical results.
The critical line for the superconducting transition is shown to coincide
with the analytical factorization line identifying the commensurate-incommensurate
transition in the XY model. 
\end{abstract}

\pacs{71.10.Hf, 75.10.Pq, 71.10.Fd}

\maketitle
The Hubbard Hamiltonian and its extensions are known to model several
correlated quantum systems, ranging from high-$T_{c}$ superconductors
to cold fermionic atoms trapped into optical lattices \cite{BDZ}.
In particular, the HBC model describes the interaction between fermions
located on bonds and on lattice sites \cite{HBC,HBC-rev}. This extension
is considered to be especially relevant to the field of high-$T_{c}$
superconductors \cite{HIRSCH}. In fact, it has recently been found
\cite{ADMO,AAA} that a superconducting phase takes place also for
repulsive values of the on-site Coulomb interaction. The phase is
characterized by incommensurate modulations in the charge structure
factor. Its boundaries have been explored numerically, though their
fundamental nature has not been understood yet. 

We find that the explanation of the above features resides into the
underlying effective model, which for the superconducting phase turns
out to be the anisotropic XY chain in a transverse field. Such model
is known to be equivalent to free spinless fermions and it is remarkable
how it can faithfully describe quantities of a strongly correlated
system like the HBC chain. Indeed, the mapping allows us to derive
analytical expressions for both the critical line and correlations,
reproducing with amazing accuracy the numerical data.

The model Hamiltonian for the HBC chain reads

\begin{align}
\mathcal{H}= & -\sum_{i\sigma}\left[1-X\left(n_{i\bar{\sigma}}+n_{i+1\bar{\sigma}}\right)\right](c_{i\sigma}^{\dagger}c_{i+1\sigma}+c_{i+1\sigma}^{\dagger}c_{i\sigma})\nonumber \\
 & +U\sum_{i}n_{i\uparrow}n_{i\downarrow}-\frac{U}{2}\sum_{i\sigma}n_{i\sigma}\label{eq:Hirsch}\end{align}
 where $\sigma=\uparrow,\downarrow$ ($\bar{\sigma}$ denoting the
opposite of $\sigma$), and the operator $c_{i\sigma}^{\dagger}$
creates a fermion at site $i$ with spin $\sigma$. Moreover $n_{i\sigma}=c_{i\sigma}^{\dagger}c_{i\sigma}$.
The parameters $U$ and $X,$ expressed in units of the hopping amplitude,
are the on-site and bond-charge Coulomb repulsion respectively. 

While the HBC model cannot be exactly solved for all $X$, there are
two integrable point at $X=0$ and $X=1$, for all values of $U$.
The former is the well-known Hubbard model which is solvable by Bethe
Ansatz. The integrability of the case $X=1$ is due to the fact that
the empty and the doubly occupied sites in this case are indistinguishable,
and the same holds for the $\uparrow$ and $\downarrow$ spins in
the singly occupied sites, so that the model can be rephrased in terms
of tight-binding spinless fermions in 1D \cite{AA}. In addition,
the number of double occupancies turns out to be a conserved quantity.

In the general case, Eq.(\ref{eq:Hirsch}) can be fruitfully recasted
passing to a slave boson representation. One can make the transformation
$|0\rangle\to e_{i}|0\rangle$, $c_{i\sigma}^{\dagger}|0\rangle\to f_{i\sigma}^{\dagger}|0\rangle$
and $c_{i\uparrow}^{\dagger}c_{i\downarrow}^{\dagger}|0\rangle\to d_{i}|0\rangle$,
where empty and doubly occupies sites are bosons, while the single
occupations are fermions. The hard-core constraint $e_{i}^{\dagger}e_{i}+d_{i}^{\dagger}d_{i}+\sum_{\sigma}f_{i\sigma}^{\dagger}f_{i\sigma}=1$
completes the identification. Then, the $c$-fermions are $c_{i\sigma}^{\dagger}=f_{i\sigma}^{\dagger}e_{i}+d_{i}^{\dagger}f_{i\bar{\sigma}}$
and $n_{i\sigma}=c_{i\sigma}^{\dagger}c_{i\sigma}=f_{i\sigma}^{\dagger}f_{i\sigma}+d_{i}^{\dagger}d_{i}$.
The total number of particles is $N=N_{f}+2N_{d}$. The filling factor
is $\nu=N/L$, with $0\leq\nu\leq2$. Accordingly, we have $\nu_{e}+\nu_{f}+\nu_{d}=1$
and $\nu=\nu_{f}+2\nu_{d}$. After the substitution, the Hamiltonian
becomes $\mathcal{H}=\sum_{i\sigma}\mathcal{H}_{i\sigma}$, where
\begin{align}
\mathcal{H}_{i\sigma} & =-\frac{U}{2}f_{i\sigma}^{\dagger}f_{i\sigma}+\left[f_{i\sigma}^{\dagger}f_{i+1,\sigma}\left(t_{X}d_{i+1}^{\dagger}d_{i}-e_{i+1}^{\dagger}e_{i}\right)\right.\nonumber \\
 & \left.-s_{X}f_{i\sigma}^{\dagger}f_{i+1,\bar{\sigma}}^{\dagger}\left(e_{i+1}d_{i}+d_{i+1}e_{i}\right)+\mathrm{H.c.}\right],\label{eq:H_slave}\end{align}
 with $t_{X}=1-2X$, and $s_{X}=1-X$. It can be recognized that the
first two terms describe the kinetic energy of a single electron (hole)
with spin $\sigma$ in a background of empty (doubly occupied) sites,
whereas the third term describes the transformation of two opposite
spins into an empty and a doubly occupied site. Since the coefficient
$s_{X}$ turns out to give the smallest contribution for $X>2/3$,
it is not surprising that the exact solution obtained assuming $s_{X}=0\mbox{ }$
(and arbitrary $t_{X})$ \cite{MON} shares in this regime many features
of the ground state of the true model, obtained by numerical investigation
\cite{ADM2009}. To some extent, these features hold within the range
$X>X_{c}=1/2$, where $X_{c}$ is the value at which $t_{X}$ changes
sign. This is true in particular as for the presence of phase coexistence
of domains formed by only empty or doubly occipied sites, in which
the single particles move. On the other hand, fixing $s_{X}=0$ yields
to a critical curve $U_{PS}=4X$ for the stabilility of the phase
separated region, whereas the superconducting transition takes place
(only for $s_{X}\neq0$) at a value $U_{SC}$ which is well below
that line. Since the role of empty and doubly occupied sites, as well
as the conservation of their number, appears to be the same as for
$s_{X}=0$ \cite{ABM} also in the superconducting case, one can infer
that it is just the motion of the single electrons and holes which
determines the change $U_{PS}\rightarrow U_{SC}$ for $s_{X}\neq0$.
In this paper, we assume this point of view: \emph{treating the empty
and doubly occupied states as the vacuum in which the single particles
move}.

Let us go back to Eq.(\ref{eq:H_slave}) and consider what happens
at $s_{X}\neq0$. The $SU(2)$ charge symmetry is broken down to $U(1)$,
which merely describes the conservation of the number of fermions.
The large spin degeneracy is removed and it is like as if the fermionic
dynamics is influenced by the background imposed by the bosons and
viceversa. This picture is correct as far as the spin and charge degrees
of freedom are not separated. For $X\lesssim1$, the pair creation
term in Eq.(\ref{eq:H_slave}) induces short-ranged antiferromagnetic
(AFM) correlations in both spin and pseudospin degrees of freedom.
Since at half-filling the probabilities of having an empty and a doubly
occupied sites are identical and coincide with 1/2, we can approximate
the term $\langle e_{i+1}^{\dagger}e_{i}-t_{X}d_{i+1}^{\dagger}d_{i}\rangle\approx X$.
Thus, in this case the kinetic energy term in $\mathcal{H}_{i\sigma}$
becomes $-X\, f_{i\sigma}^{\dagger}f_{i+1\sigma}+(X-1)\, f_{i\sigma}^{\dagger}f_{i+1,\bar{\sigma}}^{\dagger}+\mathrm{H.c.}$
where the term $f_{i\sigma}^{\dagger}f_{i+1,\bar{\sigma}}^{\dagger}$
always takes place due to the bosonic AFM correlations. Both the bosonic
species are considered as a unique vacuum for the fermions $f$.

Assuming the existence of AFM correlations also in the fermionic variables,
we can drop the spin indices. The effect of $f_{i}^{\dagger}f_{i+1}^{\dagger}$
is to open a gap at the Fermi level, hence reducing considerably the
ground state (GS) energy. This mechanism is analogous to what happens
in the case of the Peierls instability (in that case the gap is opened
by the dimerization) where the bosons here play the role of the phonons
that distort the lattice. So, we obtain a free-spinless fermion model
$\mathcal{H}^{(f)}=\sum_{i=1}^{L}\mathcal{H}_{i}^{(f)}$, where

\begin{align}
\mathcal{H}_{i}^{(f)}= & -X\left(f_{i}^{\dagger}f_{i+1}+\frac{1-X}{X}f_{i}^{\dagger}f_{i+1}^{\dagger}+\mathrm{H.c.}\right)-\frac{U}{2}f_{i}^{\dagger}f_{i}\:.\label{eq:Hf}\end{align}
 It is instructive to notice that even in this form one can recover
the exact solution of the case $X=1$. Indeed a straghtforward diagonalization
in Fourier space gives $H=-2\sum_{k}[\cos k+U/4]f_{k}^{\dagger}f_{k}$.
The fermions fill the negative energy states up to the Fermi point
$k_{f}=\pi\nu_{f}$. The saturation occurs for $U_{c}=-4\cos(\pi\nu)$
for $0<\nu<2$.

In the general case, $\mathcal{H}^{(f)}$ can be easily shown to be
equivalent to the following XY model in a transverse field\begin{equation}
\mathcal{H}_{XY}=E_{0}-\frac{1}{\zeta}\sum_{i=1}^{L}\left[\frac{1+\gamma}{2}\sigma_{i}^{x}\sigma_{i+1}^{x}+\frac{1-\gamma}{2}\sigma_{i}^{y}\sigma_{i+1}^{y}+h\sigma_{i}^{z}\right]\label{eq:Ising_map}\end{equation}
 where $\gamma=\frac{1-X}{X}$, $h=\frac{U}{4X}$, $E_{0}=-\frac{UL}{4}$
and $\zeta=\frac{1}{X}$, at half filling. As usual we have applied
the Jordan-Wigner transformation $\sigma_{i}^{z}=2f_{i}^{\dagger}f_{i}-\mathbb{I}$,
$\sigma_{i}^{+}=f_{i}^{\dagger}K_{i-1}$, $\sigma_{i}^{-}=K_{i-1}^{\dagger}f_{i}$
with $K_{l}=\prod_{k=1}^{l}\left(-\sigma_{k}^{z}\right)=\exp[i\pi\sum_{k=1}^{l}n_{k}]$.

AFM correlations in both bosonic and fermionic particles are here
assumed on the intuitive basic observation of the reduction of GS
energy by means of the pair creation terms. A more rigorous approach
would involve a self-consistent determination of the hopping coefficients
in the quadratic model in which the spin labels are retained. Such
approach allows to extend the analysis away from half filling and
in magnetic field, and goes beyond the purpose of the present paper.
We dedicate a forthcoming extended manuscript to a self-consistent
approach.

In what follows, we examine some important consequences that can be
derived from the exact solution of the XY model, written in Eq.(\ref{eq:Ising_map}).

As known, the Hamiltonian (\ref{eq:Ising_map}) can be diagonalized:
$H=E_{0}+\frac{1}{\zeta}\sum_{k\in BZ}\Lambda_{k}\left(\beta_{k}^{\dagger}\beta_{k}-\frac{1}{2}\right),$
where the sum is performed in the Brillouin zone (BZ), and the dispersion
relations are $\Lambda_{k}=2\sqrt{\left(\cos k+h\right)^{2}+\gamma^{2}\sin^{2}k}$.
Given the positiveness of $\Lambda_{k}$, the GS energy $E_{GS}$
is determined by the vacuum of the Bogoliubov quasiparticles $\beta_{k}$,
giving $E_{GS}=E_{0}-\frac{1}{2L}\sum_{k\in BZ}\Lambda_{k}$. By taking
the thermodynamic limit $L\to\infty$, we get an energy density $e_{GS}=-\frac{U}{4}-\frac{X}{4\pi}\int_{-\pi}^{\pi}dk\Lambda_{k}$.

We have compared the outcomes of our mapping with numerical calculations
using the density matrix renormalization group (DMRG) \cite{S2005}.
In particular, we used extrapolations in $1/L$ of data collected
by selecting 7 finite-system sweeps and 1024-1152 states. Numerical
and analytical results of the energy density at $X=0.8$ are displayed
in table \ref{tab:compx08}. %
{}

\begin{table}[th]
\begin{tabular}{|c|c|c|c|c|c|c|}
\hline 
$U$  & $e_{GS}^{num}$  & $e_{GS}^{th}$  & $\nu_{d}^{num}$  & $\nu_{d}^{th}$  & $q/\pi$  & $\psi/\pi$\tabularnewline
\hline
\hline 
$0$  & $-0.5390$ & $-0.54612$ & $0.2511$ & $1/4$ & $14/30$ & $1/2$\tabularnewline
\hline 
$0.5$  & $-0.670$ & $-0.67708$ & $0.216$ & $0.22611$ & $14/30$ & $0.44841$\tabularnewline
\hline 
$1$  & $-0.81544$ & $-0.82011$ & $0.19016$ & $0.20164$ & $12/30$ & $0.39539$\tabularnewline
\hline 
$1.5$  & $-0.9717$ & $-0.97565$ & $0.173$ & $0.17588$ & $10/30$ & $0.33914$\tabularnewline
\hline 
$2.5$  & $-1.3300$ & $-1.3287$  & $0.1063$ & $0.11488$ & $6/30$ & $0.20116$\tabularnewline
\hline
\end{tabular}

\caption{Comparison between various quantities defined in the text computed
either numerically (num) with DMRG or analytically by means of the
equivalent XY model (th) for $X=0.8$, both for periodic boundary
conditions. The latter is treated directly in the thermodynamic limit,
while the former are extrapolated to $L\to\infty$ from finite-size
data. The characteristic wavenumber $q$ is extracted from Fourier
transforms at $L=30$.\label{tab:compx08}}

\end{table}

An important feature of the XY chain is the presence of a factorization
line $h^{2}+\gamma^{2}=1$, which corresponds to a commensurate-incommensurate
(CIC) transition. In the HBC model this transition is mapped analytically
into \begin{equation}
U_{SC}=4\sqrt{2X-1}.\label{Usc}\end{equation}

\begin{figure}
\includegraphics[scale=0.27]{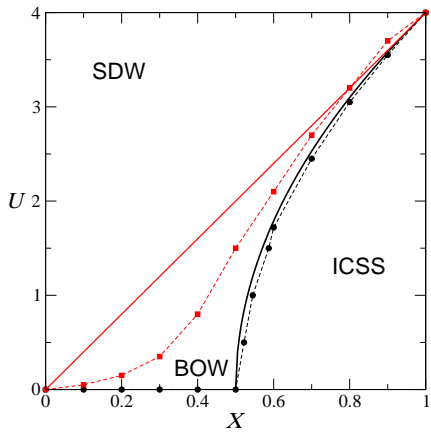}

\caption{Comparison between the phase diagram of the HBC chain, calculated
numerically in Ref.\cite{AAA} (symbols with dashed lines) and the
phase diagram obtained from the mapping onto the XY model in transverse
field (continuous lines). The upper curves correspond to the spin
gap transition where spin excitations become gapless, while the lower
curves mark the transition into the ICSS phase where the charge compressibility
diverges.\label{fig:pdhf}}

\end{figure}

Such transition was discovered numerically in Ref.\cite{ADMO} and
separates a incommensurate singlet superconducting (ICSS) phase from
a bond ordered wave (BOW) phase \cite{AAA}. As seen in Fig.\ref{fig:pdhf},
the curve obtained with our mapping describes rather accurately the
numerical data of the transition. %

Along the factorization line the GS in the $S=1/2$ model is written
as $\otimes_{i=1}^{L}|\phi\rangle$, where $|\phi\rangle=\cos\frac{\theta}{2}|\uparrow\rangle+\sin\frac{\theta}{2}|\downarrow\rangle$,
with $\cos\theta=[(1-\gamma)/(1+\gamma)]^{1/2}=\alpha.$ Here the
local magnetization is $2\nu_{f}-1=\alpha=\sqrt{2X-1}.$ Accordingly,
the number of double occupations along the factorization line at half
filling is $\nu_{d}=(1-\alpha)/4=(1-\sqrt{2X-1})/4.$ In the rest
of the phase diagram, the transverse magnetization of the XY chain
is given by $\langle\sigma_{i}^{z}\rangle=\frac{2}{L}\sum_{k\in BZ}(h+\cos k)\Lambda_{k}^{-1}$.
The diverging charge compressibility of the ICSS phase is explainable
simply by observing that adding two particles produce the conversion
of an empty site onto a doubly occopied one, without changing the
energy in the XY representation. 

In addition, the XY model in 1D is known to undergo a quantum phase
transition along line $h=1$, belonging to the universality class
of the classical Ising model in 2D. This translates directly into
the line $U=4X$ in the phase diagram of the HBC model (see Fig.\ref{fig:pdhf}).
The latter coincides with the critical line of stability of PS ($U_{PS}$)
in the integrable case $s_{X}=0$, and is close to the numerical critical
line between the spin density wave (SDW) and the BOW phase in Fig.
1, at least for $X$ close to 1. Moreover, the line $\gamma=1$, which
is known to describe the Ising model in a transverse field, here corresponds
to the case $X=1/2$. While it is questionable whether the assumptions
that have originated our approximations for the ICSS phase are still
valid in the above limiting cases, one can recognize that instead
at the very crucial critical point $X=1/2$ and $U=0$, our system
described in Eq.(\ref{eq:Hirsch}) is mapped into nothing but the
Ising model. 

From the seminal paper of Barouch and McCoy on the statistical mechanics
of the XY model \cite{BM1971}, it is known that the oscillation wavenumber
of the correlator $\rho_{xx}(R)=\langle\sigma_{i}^{x}\sigma_{i+R}^{x}\rangle$
in the incommensurate region $h^{2}+\gamma^{2}<1$ is \begin{align}
\tan\psi & =\frac{\sqrt{1-\gamma^{2}-h^{2}}}{h}=\sqrt{\frac{2X-1}{\left(U/4\right)^{2}}-1},\label{eq:peak}\end{align}
 with a period $R_{0}=2\pi/\psi$.

\begin{figure}
\includegraphics[scale=0.26]{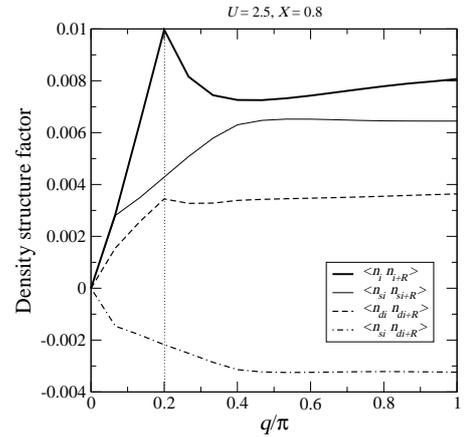}

\caption{Analysis of the various contributions {[}see Eq.(\ref{eq:contnn}){]}
to the static structure factor (Fourier transform) of the density
correlation function for the HBC model ($L=30)$ with the parameters
reported in the legend. The vertical line corresponds to $\psi/\pi$
(see table \ref{tab:compx08}).\label{fig:FTnntds}}

\end{figure}

A first striking observation is the fact that the correlations of
the \textit{total} density exhibit a peak very close to the characteristic
wavenumber $\psi$ in Eq.(\ref{eq:peak}): in the last two columns
of table \ref{tab:compx08} we report the wavevector $q$ at which
the total density structure factor has a peak (see an example in fig.
\ref{fig:FTnntds}) and the corresponding value of $\psi$. For $X=0.9$
and $U=3$ with $L=32$ the peak is located at $q/\pi=0.1875$ while
$\psi/\pi=0.18342$.

The appearance of the peak at wavenumber $Q$ in the Fourier transform
of a correlation function that decays as $\cos(QR)\exp(-R/\xi)/R^{a}$
is related also to the exponent $a$: the smaller is $a$ the sharper
is the peak. In particular for $a=2$ which is the case for the correlation
$\rho_{zz}(R)=\langle\sigma_{i}^{z}\sigma_{i+R}^{z}\rangle$ of the
XY model, the peak it not visibile at all, despite the fact that the
oscillations actually \textit{\textcolor{black}{have}} characteristic
wavenumber $2\psi$. Hence, it is worth to inspect in more detail
the origin of the peaks observed numerically. Since the local density
operator $n_{i}$ in terms of single and double occupancies $n_{si}$
and $n_{di}$ is given by $n_{i}=n_{si}+2n_{di}$, the correlation
function of the total density decomposes in the following parts\begin{align}
\langle n_{i}n_{i+R}\rangle= & \langle n_{si}n_{si+R}\rangle+4\langle n_{di}n_{di+R}\rangle\nonumber \\
 & +2\langle n_{di}n_{si+R}\rangle+2\langle n_{si}n_{di+R}\rangle.\label{eq:contnn}\end{align}
 It turns out that the peak in the static structure factor is not
due to the first term, but it is instead provided by $\langle n_{di}n_{di+R}\rangle$,
as shown in Fig.\ref{fig:FTnntds} for the test case $X=0.8$ and
$U=2.5$, although we obtained the same qualitative picture at $U=1$.

\begin{figure}
\includegraphics[scale=0.27]{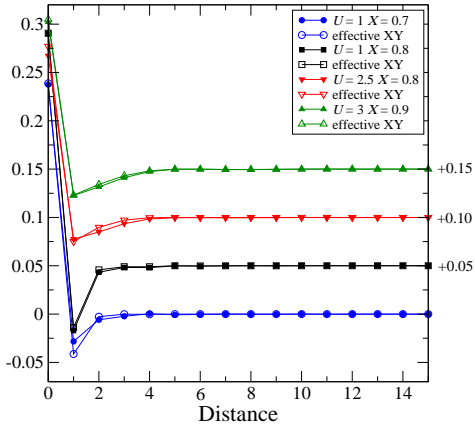}

\caption{Comparison of the connected real-space correlation functions $N_{s}(R)$
(at half-filling) and $\rho(R)$ for the HBC and XY model, respectively
(see text for definitions based on singly occupied sites operators).
The parameters of the two models are related by mapping as $\gamma=(1-X)/X$
and $h=U/4X$. All the DMRG calculations for the HBC model and the
analytical experessions of the curves for the XY model refer to $L=50$.
From top to bottom the data have been offset by +0.15, +0.10 and +0.05
for the sake of clarity.\label{fig:corrs_Hirsch_vs_XY}}

\end{figure}

According to our mapping, we can compare directly the connected correlator
$N_{s}(R)=\langle n_{si}n_{si+R}\rangle-\langle n_{si}\rangle^{2}$
in the HBC model with the density correlation function $\rho(R)=\langle f_{i}^{\dagger}f_{i}f_{i+R}^{\dagger}f_{i+R}\rangle-\langle f_{i}^{\dagger}f_{i}\rangle^{2}$
for the spinless fermions with Hamiltonian (\ref{eq:Hf}). The calculation
of the latter is omitted here since it is quite lengthy, though it
simply involves a standard application the Wick theorem. The fully
fermionic correlator $N_{s}(R)$ and the spinless fermions correlator
$\rho(R)$ are compared in Fig.\ref{fig:corrs_Hirsch_vs_XY} for various
choices of the parameters $U$ and $X$ in the ICSS phase of our starting
system (i.e. the incommensurate one in the XY model); the agreement
in real space is generally very good. Such behaviour of $N_{s}(R)$
is not obvious a priori in the HBC model and we interpret it as a
remarkable nontrivial prediction of our mapping.

In summary, we have studied the Hubbard model with bond-charge interaction
in the superconducting regime, unveiling its underlying XY structure.
We have shown that at half filling the numerical critical line for
superconductivity coincides with remarcable accuracy to the analytical
factorization curve that marks the CIC transition of the anisotropc
XY model in a transverse field. Exploting the mapping for the calculation
of correlations in the effective model has allowed us to predict rather
accurately the peak in the charge structure factor of the original
model. The results confirm a posteriori the crucial role of short
range AF correlations and spin degrees of freedom as to the onset
of superconductivity. The ultimate presence of the latter is however
to be ascribed to the interplay of the spin with the charge degrees
of freedom, the superconducting properties being absent from the incommensurate
phase of the free fermions model.

Based on the success of the present mapping, a number of further result
are now in order. First, since the one-dimensionality of the model
is not crucial to the mapping, the latter should hold in higher dimension
as well. In 2D, the numerical investigation of the XY model has been
largely explored in the literature: this could provide useful hints
on the type of phase diagram which characterizes the 2D HBC model.
Moreover, it would be interesting to understand the implications on
the HBC model of a non-vanishing string order parameter which is peculiar
of the XY model in transverse field. Finally, we expect that a similar
mapping should hold also in the strongly repulsive regime $U\rightarrow\infty,$
since in that case no doubly occupied sites occur, and it is still
quite natural to assume short range AFM order of single particles. 
\begin{acknowledgments}
We are grateful to Alberto Anfossi for useful discussions, and for
providing us some data to compare. AM acknowledges the hospitality
of Condensed Matter Theory Visitor's Program at Boston University,
where this work was completed. The Bologna Section of the INFN is
also acknowledged for the computational resources. This work was partiall
supported by national italian funds, PRIN2007JHLPEZ\_005. \end{acknowledgments}

\end{document}